\documentclass[final,1p]{elsarticle}
\usepackage{booktabs}
\usepackage{multirow}
\usepackage[table,xcdraw]{xcolor}
\usepackage{caption}

\usepackage{amssymb}

\usepackage{xcolor}
\journal{TBD}

\begin{document}

\begin{frontmatter}

\title{A Reactive Force Field Approach to Modeling Corrosion of NiCr Alloys in Molten FLiNaK Salts}

\author[1]{Hamdy Arkoub}
\author[2]{Swarit Dwivedi}
\author[2]{Adri C.T. van Duin}
\author[1]{Miaomiao Jin\corref{cor1}}
\cortext[cor1]{Corresponding: M. Jin (mmjin@psu.edu)}
\affiliation[1]{organization={Department of Nuclear Engineering, The Pennsylvania State University},
            city={University Park},
            postcode={16802}, 
            state={PA},
            country={USA}}
            
\affiliation[2]{organization={Department of Mechanical Engineering, The Pennsylvania State University},
            city={University Park},
            postcode={16802}, 
            state={PA},
            country={USA}}

\begin{abstract}
The interface between NiCr alloys and FLiNaK molten salt exhibits complex corrosion behavior, mainly driven by intricate chemical interactions involving Cr and F$\mathrm{^-}$ ions. Understanding these dynamic reactions is crucial for developing effective corrosion mitigation strategies to ensure the long-term durability of Ni-based alloy components in molten salt technologies. However, obtaining molecular-level understanding through experiments is challenging. To address this, we utilize reactive molecular dynamics simulations enabled by a reactive force field, ReaxFF, to investigate detailed reaction dynamics at the atomic level. Since there is currently no available force field involving fluoride salt and Ni-based alloys, we first present the development of the ReaxFF parameter set for Ni/Cr/F/Li/Na/K based on extensive first-principles calculations. With this force field, we achieve a strong agreement for the structure of FLiNaK molten salt by comparing the pair distribution functions with experimental and simulation results. Furthermore, it successfully reproduces the experimental phenomenon of Cr dissolution in fluoride salt, with the corrosion rate depending on the alloy and salt compositions. Particularly, it reveals that increasing the concentration of Li can enhance the formation of a compact double layer, mitigating Cr dissolution.  This work enables a fundamental understanding of the interfacial behavior between fluoride salt and NiCr alloys.  

\end{abstract}


\end{frontmatter}


\section{Introduction}
The molten-salt reactor (MSR) has emerged as a promising next-generation nuclear reactor due to its numerous advantages, such as improved efficiency, online refueling, and minimal nuclear waste production \cite{leblanc2010molten,kelly2014generation}. Molten fluoride salts such as FLiBe and FLiNaK are the key candidates of the coolant and/or fuel carriers, given their favorable chemical stability and thermophysical characteristics \cite{williams2006assessment,delpech2010molten}. However, at high temperatures ($> \mathrm{700 \,^o C}$), these highly corrosive molten salts pose significant challenges to structural materials. Unlike aqueous environments where corrosion mitigation relies on a passivating layer of oxide scale on the metal surface, the oxide layer is unstable in molten salts \cite{sohal2010engineering}. Furthermore, salt corrosivity is enhanced due to various impurities such as fission products, moisture, and polyvalent metallic ions (e.g., Ni$\mathrm{^{2+}}$, Cr$\mathrm{^{3+}}$, Fe$\mathrm{^{2+}}$) \cite{calderoni2012corrosion,ouyang2013effect,yin2018effect,wang2016effects}. Moreover, corrosion can be escalated due to thermal gradients and dissimilar materials \cite{wang2014galvanic,keiser1977compatibility,zheng2015corrosion,olson2009materials}. Thermodynamically, Cr would be a more susceptible alloy element compared to other metal elements such as Mo, Ni, and Fe due to a more negative free energy of reaction \cite{olson2009materials,mcalpine2020corrosion,devan1962corrosion}. Therefore, degradation of the Cr oxide layer and selective dissolution of Cr from the bare metals \cite{olson2009materials,zhou2020proton,olson2010intergranular,sridharan2013corrosion} deem Cr-rich alloy unusable. 

Ni-based alloys have been identified as the top candidate materials for MSR structural components due to their favorable resistance to fluoride salt corrosion, high-temperature creep, and radiation damage \cite{yvon2009structural,rosenthal1972development}. Despite their general resilience, studies have shown that corrosion with Cr dissolution from the alloys is dominant in molten fluoride salt environments, and with the degree of corrosion closely tied to the Cr concentration in the alloys \cite{sohal2010engineering,williams2006assessment,olson2009materials,ouyang2013effect,leong2023kinetics}. The corrosion process involves intricate chemical reactions occurring at the alloy-salt interface, where corrosive species in molten salts react with active alloying elements, particularly Cr, which diffuses from the bulk to the surface of the alloys \cite{olson2009materials}. Previous studies highlight the significant role of F in the corrosion of Ni-based alloys in fluoride molten salts, as there is strong F-Cr bonding leading to weakened Ni-Cr bonding \cite{ren2016adsorption}. Furthermore, it was shown that with greater F coverage, extensive F chemisorption can significantly alter the surface morphology of the alloy, leading to the selective dissolution of Cr in the form of $\mathrm{CrF_2}$\slash $\mathrm{CrF_3}$ molecules in the salt \cite{yin2018first,chan2022insights}. Cr$^{3+}$ has been shown to be the primary charge state of Cr in salt \cite{williams2006assessment}.  Meanwhile, the strong F-Cr interaction enhances Cr segregation, although this impact can be screened and diminished by adjacent cations \cite{startt2021ab}. Even with numerous work on this topic, to the best knowledge of the authors, there is still a lack of understanding of how interfacial dynamics (particularly Cr dissolution) would be affected by local atomic environment arising from varying alloy and salt compositions. 

Modeling the early-stage corrosion mechanisms of Ni-based alloys in molten fluoride salts has relied on first-principles methods such as density functional theory (DFT). However, these methods are limited to systems of a few hundred atoms and a timescale of a few picoseconds due to prohibitive computational costs. This limitation hinders a comprehensive understanding of dynamic processes and overall material behavior. As an alternative, reactive molecular dynamics (RMD) based on reactive force fields like ReaxFF offer a balance between computational cost and accuracy to capture the dynamic processes \cite{senftle2016reaxff}. RMD simulations of the interfacial region can expand the length scale to microns and capture the surface transport beyond DFT calculations. Here, we utilize RMD with the ReaxFF force field \cite{senftle2016reaxff,van2001reaxff} to capture the surface processes. ReaxFF has been used frequently to study interfacial chemistry, for example, aqueous corrosion of Cu \cite{jeon2011atomistic}, Ni \cite{assowe2012reactive}, and Fe \cite{dormohammadi2019investigation}. RexaFF parameters are transferable within the same development branch, enabling merging of simulations of reactive events between solid and liquid phases \cite{senftle2016reaxff,van2001reaxff}. The initial corrosion of NiCr alloys in the fluoride salt fits into this method to reveal the active unit mechanisms. 

In this study, we develop a ReaxFF parameter set for the Ni/Cr/F/Li/Na/K system based on extensive first-principles calculations. This parameter set enables a fundamental understanding of the interfacial behavior between fluoride salt and NiCr alloys. By varying the salt and alloy composition separately, we aim to shed light on the major factors influencing the corrosion of Ni-based alloys in molten fluoride salts. 

\section{Methods}

\subsection{ReaxFF Force Field}

The ReaxFF method using a bond-order formalism bridges the gap between computationally intense DFT calculations and classical MD without reactive properties. It describes many-body interactions with various potential energy terms, with the total energy of a system written as,

\[E{\mathrm{_{total}}}=E{\mathrm{_{bond}}}+E{\mathrm{_{over}}}+E{\mathrm{_{tors}}}+E{\mathrm{_{angle}}}+E{\mathrm{_{VdW}}}+E{\mathrm{_{Coul}}}+E{\mathrm{_{Specific}}}\]

where ${E{\mathrm{_{bond}}}}$, $E{\mathrm{_{over}}}$, $E{\mathrm{_{tors}}}$, $E{\mathrm{_{angle}}}$, $E{\mathrm{_{VdW}}}$, and $E{\mathrm{_{Coul}}}$ refer to the bond order energy, the over-coordination penalty energy, the torsion angle energy, the three-body valence angle strain, van der Waals interactions, and the coulomb interactions, respectively. $E{\mathrm{_{Specific}}}$ includes other necessary properties to the system of interest, like lone-pair and conjugation. These energy expressions depend on the bond order (BO$_{ij}$) between atoms $i$ and $j$, which is influenced by the interatomic distances and the local environment. For the sake of brevity, readers are referred to these references \cite{van2001reaxff,chenoweth2008reaxff,senftle2016reaxff} for detailed formulation and physical interpretation of each term. The unknown parameters involved in each term need to be trained against first-principles calculations. Overall, general parameters, atom parameters, bond parameters, off-diagonal parameters, angle parameters, torsion parameters, and hydrogen-bond parameters make up a typical set of ReaxFF parameters. Given that there are six elements involved, to constrain the problem within a reasonable computational expense, parameters for certain atom types are taken from previous work to be used in the current force field, or as an initial guess for further training. 

The parameters for Ni and Cr are obtained from Shin et al. \cite{shin2021impact}, which have shown good performance in predicting structural, mechanical and thermal properties for Ni, Cr, and their alloys. The F atom parameters and F-F bonded and non-bonded parameters are taken from the work by Liu et al. \cite{liu2019formation}. For the development of the missing F-Ni/Cr parameters (bond, off-diagonal, angle, and torsion), Al-F parameters from Liu et al. \cite{liu2019formation} are taken as an initial guess. This force field parameter set of Ni/Cr/F atoms is optimized against the DFT training data containing various F-Ni/Cr interactions (Section \ref{sec:dft}). In the FLiNaK salt, the F-K parameters are adopted from the work of Fedkin et al. \cite{fedkin2019development} on the water electrolyte system including alkali metal cations (Li$\mathrm{^+}$, Na$\mathrm{^+}$, K$\mathrm{^+}$, Cs$\mathrm{^+}$) and halogen anions (F$\mathrm{^-}$, Cl$\mathrm{^-}$, and I$\mathrm{^-}$). Then, these parameters were used as initial guesses for F-Li and F-Na, which are then trained against DFT data (Section \ref{sec:dft}). For the salt cation interactions, the parameters are obtained from the work by Fedkin et al. \cite{fedkin2019development}. Based on the assumption that the FLiNaK/NiCr interface chemistry will mainly involve F-binding to the NiCr surface, the bond parameters for Ni/Cr-salt cations are not trained in this work and were treated as dummy bonds. For validation of this potential, we compare both energetics and dynamical properties with DFT calculations and experimental data if available.

\subsection{First Principles Calculations}
\label{sec:dft}
Various quantities are computed based on DFT as implemented in the Vienna Ab initio Simulation Package (VASP) \cite{kresse1993vasp,kresse1996efficiency,kresse1996efficient} and Quantum Espresso package \cite{giannozzi2009quantum,giannozzi2017advanced} to build the reference values for the training set. This set consists of gas-phase calculations, the heat of formation energies, F adsorption energies, F interstitial formation energies, and equations of state. The Ni 4s$\mathrm{^2}$3d$\mathrm{^8}$, Cr 4s$\mathrm{^1}$3d$\mathrm{^5}$, Li 2s$\mathrm{^1}$, Na 3s$\mathrm{^1}$, and F 2s$\mathrm{^2}$2p$\mathrm{^5}$ valence electrons were explicitly considered. All calculations apply spin-polarization, the projector-augmented wave (PAW) method to approximate core electrons, the Perdew-Burke-Ernzerhof \cite{perdew1996generalized} (PBE) functional within the generalized gradient approximation (GGA) to account for electron exchange correlation. More details for the following calculations are provided in the Supplementary Materials (SM). 

\textit{Gas-Phase calculations}. The DFT calculations of the bond lengths and valence angles of various gas-phase structures that can potentially form as Cr/Ni dissolves in the fluoride salt are performed. These structures include F$\mathrm{_2}$, CrF, CrF$\mathrm{_2}$, CrF$_3$, CrF$\mathrm{_4}$, CrF$\mathrm{_5}$, Cr$\mathrm{_2}$F$\mathrm{_5}$, NiF, NiF$\mathrm{_2}$, and NiF$\mathrm{_3}$. The supercell size is 20 $\times$ 20 $\times$ 20 $\AA^3$, with a single gas molecule. For geometry optimization, a kinetic energy cutoff of 30 Ry for the wave function, a charge density cutoff of 180 Ry, an energy threshold of 10$^{-6}$ Ry and the atomic force threshod of 0.00038 Ry/a.u, are applied, with a single gamma k-point. These cutoff values are determined after convergence test.

\textit{Heat of formation}. Formation energies for various metal-fluoride crystalline structures are considered. These crystal structures include NiF$_2$ (tetragonal structure), NiF$_3$ (trigonal structure), CrF$_2$ (monoclinic structure), CrF$_3$ (trigonal structure), CrF$_4$ (monoclinic structure), and CrF$_5$ (orthorhombic structure). The Heat of formation for all these structures was taken from the Materials Project Database \cite{jain2013commentary}.

\textit{F adsorption energies}. F adsorption energies at the different adsorption sites on face-centered cubic (FCC) Ni slabs, and Cr-doped FCC Ni slabs are calculated. See SM for detailed structures. The considered surface orientations include (111), (110), and (100) planes. For geometry optimization, the kinetic energy cutoff for wave function was 25 Ry, and 250 Ry for the charge density cutoff. The atomic positions are optimized until the atomic forces below 0.001 Ry/a.u. The energy convergence threshold is set to be 10$^{-3}$ Ry. A 2 $\times$ 2 $\times$ 1 k-point mesh is used. The adsorption energy ($\mathrm{E_{ads}}$) is calculated by,
\[E_{\mathrm{ads}}=E_{\mathrm{F/surface}}-E_{\mathrm{surface}}-0.5 \times E_{\mathrm{F_{2}}}\]
where $E_{\mathrm{F_{2}}}$, $E_{\mathrm{surface}}$, $E_{\mathrm{F/surface}}$ are the energies of the F$\mathrm{_2}$ molecule in vacuum, metal slab, and F adsorption on the metal slab, respectively.

\textit{F interstitial formation energy}. The formation energies of F interstitial at the octahedral and tetrahedral sites for both FCC Ni and body-centered cubic (BCC) Cr structures are computed. Supercells of 2 $\times$ 2 $\times$ 2 primitive cells are constructed for the two cases. For geometry optimization, Monkhorst-Pack k-point mesh of 2 $\times$ 2 $\times$ 2 and energy cutoff value of 400 eV are used for the two cases. F interstitial formation energy ($\mathrm{E_{int}}$) is calculated by,
\[E_{\mathrm{int}}=E_{\mathrm{F/lattice}}-E_{\mathrm{lattice}}-0.5 \times E_{\mathrm{F_{2}}}\]
where $E_{\mathrm{F{_2}}}$, $E_{\mathrm{lattice}}$, $E_{\mathrm{F/lattice}}$ are the energies of the F$\mathrm{_2}$ molecule in vacuum, metal lattice, and metal lattice with F at the interstitial site, respectively.

\textit{Equation of state}. The equations of state (EOS) for NiF$\mathrm{_2}$ (tetragonal structure) and CrF$\mathrm{_3}$ (trigonal structure) are considered, which possess the lowest formation energy per atom for Ni-F and Cr-F, respectively. Additionally, the EOS for LiF and NaF in rock salt structures are included. The energy convergence is set to be 0.00027 eV, with a cutoff of 400 eV. The Brillouin zone is sampled with Monkhorst-Pack k-point mesh: 3 $\times$ 3 $\times$ 4 for NiF$\mathrm{_2}$, 3 $\times$ 3 $\times$ 1 for CrF$\mathrm{_3}$, and 3 $\times$ 3 $\times$ 3 for LiF and NaF crystal structures. 
 
\subsection{ReaxFF MD simulations}

After obtaining the force field, to evaluate its capability of simulating dynamic processes, we perform three sets of RMD calculations to understand i) salt structures, ii) initial corrosion process with varying NiCr compositions, and iii) initial corrosion process with varying salt compositions. All simulations and visual representation of simulation snapshots are carried out in the AMS package \cite{van2001reaxff,chenoweth2008reaxff,reaxff2022.1scm} and the OVITO package \cite{stukowski2009visualization}. 

We first generate a box of 20 nm $\times$ 20 nm $\times$ 20 nm with 109 LiF, 28 NaF, and 99 KF molecules randomly distributed. The simulation is performed at atmospheric pressure using berendsen barostat \cite{berendsen1984molecular}, and at 600 $\mathrm{^o}$C using Nose-Hoover thermostat \cite{nose1984unified}, under periodical boundary conditions and a time step of 0.05 fs. To characterize the salt structure, the partial pair distribution functions (PDFs) of Li-F, Na-F, and K-F are calculated. The equations used to calculate the PDFs are \cite{Rüger2022AMS},
\[N(r)=\frac{1}{n_{\mathrm{frames}}} \sum_{s=1}^{n_{\mathrm{frames}}} \sum_{i=1}^{n_{\mathrm{from}}} \sum_{j=1}^{n_{\mathrm{to}}} \delta (r_{ij}^{s}-r) \]
\[g(r)=\frac{N(r)}{V(r)\rho _{tot}} \]
where $n_{\mathrm{frames}}$ is the total number of frames used to calculate the PDF, two sets of atoms $S_{\mathrm{from}}$ and $S_{to}$, have a size $n_{\mathrm{from}}$ and $n_{to}$ respectively, $r_{ij}^{s}$ is the distance between atom $i$ in $n_{\mathrm{from}}$ and atom $j$ in $n_{to}$ for trajectory frame $s$, $\rho _{tot}$ is the total density of the system, $V(r)$ is the volume of a spherical shell at radius $r$ with thickness $dr$, and $g(r)$ is the final PDF.

To simulate the corrosion process, a slab composed of 10$\times$10$\times$4 FCC unit cells (800 Ni atoms) is first generated. Here, we consider three distinct Ni/Cr compositions: Ni$\mathrm{_{0.95}}$Cr$\mathrm{_{0.05}}$, Ni$\mathrm{_{0.85}}$Cr$\mathrm{_{0.15}}$, and Ni$\mathrm{_{0.75}}$Cr$\mathrm{_{0.25}}$, hence, the alloy is generated by randomly replacing Ni atoms with Cr atoms at the specified percentage. The alloy structure is first relaxed at isothermal and isobaric condition (NPT), without the presence of salt, with a time step of 0.25 fs at 800 $\mathrm{^o}$C and atmospheric pressure for a total of 20,000 steps. Then on top of the slab, a random distribution of 163 LiF molecules, 147 KF molecules, and 40 NaF molecules (LiF–NaF–KF: 46.5–11.5–42 mol\%) is introduced; this composition is commonly seen in FLiNaK salt studies \cite{williams2006assessment}. The salt occupies a volume corresponding to a density of approximately 1.9 g/cm$^3$ at 800 $^o$C \cite{frandsen2020structure,chrenkova2003density,sohal2010engineering}. The lowermost two layers of the alloy are held static. Furthermore, to understand how the salt composition modifies the corrosion behavior, we consider Ni$\mathrm{_{0.85}}$Cr$\mathrm{_{0.15}}$ in contact with individual fluoride salt, where 300 KF molecules, NaF molecules, and LiF molecules are introduced, separately. Simulations are conducted under the constant volume and temperature condition (NVT) at 800 $\mathrm{^o}$C, with a time step of 0.25 fs. For statistical analysis, five independent simulations are performed in each scenario. 

The electronegativity equalization method (EEM) \cite{mortier1985electronegativity} is used to calculate ion charges across the simulations. This method dynamically equilibrates the atomic charges over the entire system, which can reveal the charge transfer at the interface. To evaluate F coverage on the alloy surface, we adopt a distance criterion based on the 80\% of the sum of atom Van der Waals radii, leading to $\mathrm{r_{\mathrm{cut}}\mathrm{[Cr-F]}}$= 2.776 Å and $\mathrm{r_{\mathrm{cut}}\mathrm{[Ni-F]}}$ = 2.48 Å. 

\section{Results and Discussion}
\subsection{Force Field Development and Validation}
\subsubsection{Ni\slash Cr\slash F Force Field}
To effectively capture the corrosion process facilitated by F interactions with the alloy, the initial focus is to ensure strong agreement between ReaxFF and DFT calculations regarding Ni\slash F and Cr\slash F interactions. 

\begin{figure}[!ht]
	\centering
	\includegraphics[width=0.7\textwidth]{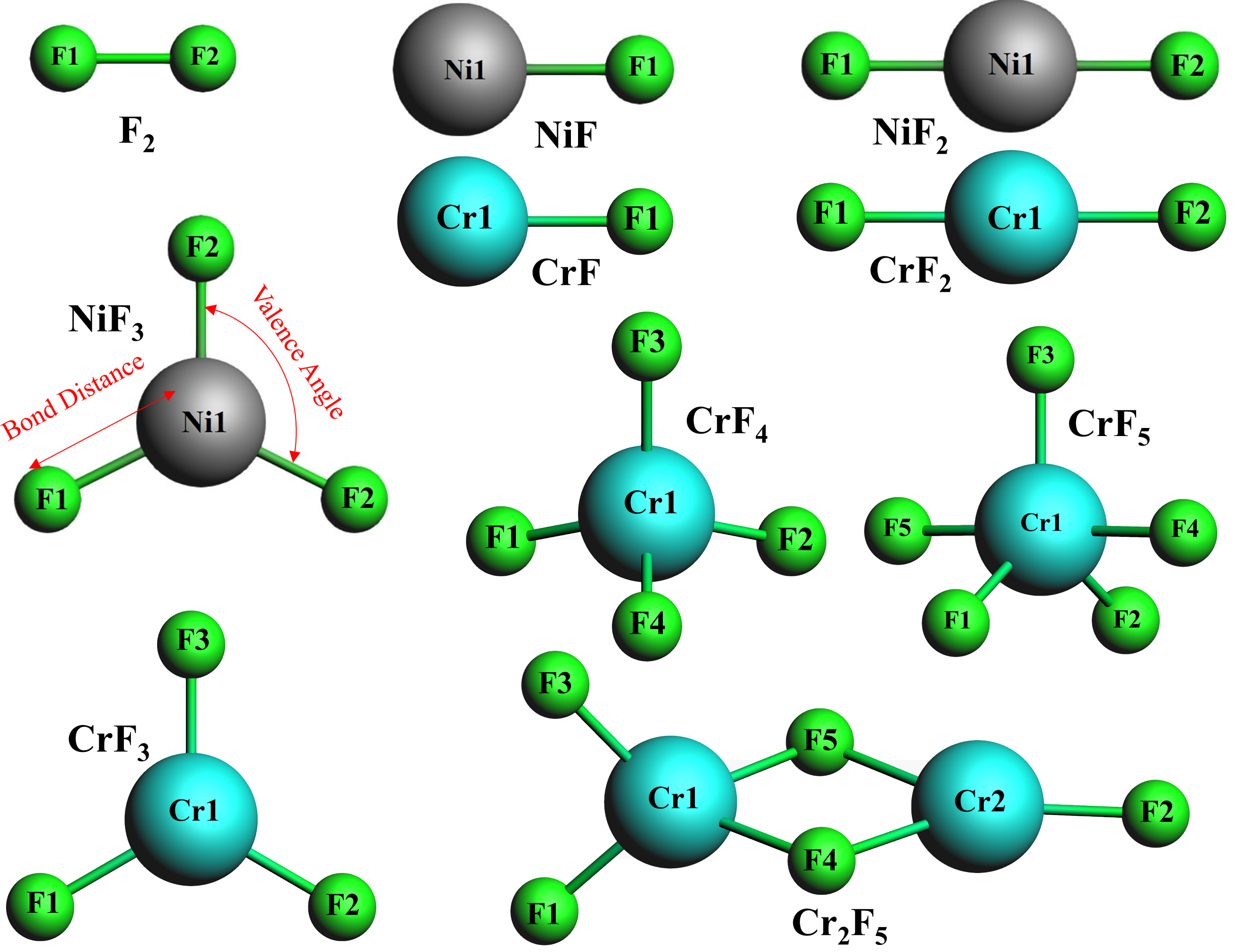}
	\caption{Gas-phase molecular structures considered in the training set.}
	\label{fig:gases}
\end{figure}

First, the molecular structures as depicted in Figure \ref{fig:gases} are considered, including $\mathrm{F_2}$, $\mathrm{CrF}$, $\mathrm{CrF_2}$, $\mathrm{CrF_3}$, $\mathrm{CrF_4}$, $\mathrm{CrF_5}$, $\mathrm{Cr_2 F_5}$, $\mathrm{NiF}$, $\mathrm{NiF_2}$, and $\mathrm{NiF_3}$. These structures have been discussed in previous studies \cite{sankar2023effect,pavlik2022accelerated,chan2022insights,li2015high,pavlik2015corrosion}. Here, we attempt to reproduce the Cr-F and Ni-F bonding, as well as the angles formed by F-Cr-F and F-Ni-F. A detailed comparison of the bond distances and valence angles based on DFT and ReaxFF calculations is presented in Tables. \ref{table_1} and \ref{table_2}, revealing a strong alignment between the two, thereby affirming the accuracy of ReaxFF in predicting bond distances and valance angles.

\begin{table}[!ht]
\footnotesize
    \centering
    \caption{Gas-phase bond distances (\AA) of various Cr-F and Ni-F structures. Refer to Figure \ref{fig:gases} for the numbering of atoms.}
    \begin{tabular}{p{1cm}p{1cm}p{1cm}p{1cm}}
    \hline
        \textbf{Formula} & \textbf{Bond} & \textbf{ReaxFF} & \textbf{DFT} \\ \hline
        \textbf{F$_2$} & 1F-2F & 1.4818 & 1.4053 \\
        \textbf{CrF} & 1Cr-1F & 1.7984 & 1.7894 \\ 
        \textbf{CrF$_2$} & 1Cr-1F & 1.8028 & 1.778 \\ 
        \textbf{} & 1Cr-2F & 1.8028 & 1.778 \\ 
        \textbf{CrF$_3$} & 1Cr-1F & 1.8005 & 1.7349 \\ 
        \textbf{} & 1Cr-2F & 1.8005 & 1.7349 \\ 
        \textbf{} & 1Cr-3F & 1.8005 & 1.7358 \\ 
        \textbf{CrF$_4$} & 1Cr-1F & 1.8061 & 1.7196 \\ 
        \textbf{} & 1Cr-2F & 1.8061 & 1.7196 \\ 
        \textbf{} & 1Cr-3F & 1.8061 & 1.7203 \\ 
        \textbf{} & 1Cr-4F & 1.8061 & 1.7202 \\ 
        \textbf{CrF$_5$} & 1Cr-1F & 1.8162 & 1.7072 \\ 
        \textbf{} & 1Cr-2F & 1.8162 & 1.7072 \\ 
        \textbf{} & 1Cr-3F & 1.8154 & 1.6944 \\ 
        \textbf{} & 1Cr-4F & 1.8412 & 1.7589 \\ 
        \textbf{} & 1Cr-5F & 1.8414 & 1.7589 \\ 
        \textbf{Cr$_2$F$_5$} & 1Cr-5F & 1.9785 & 1.9271 \\ 
        \textbf{} & 1Cr-1F & 1.8053 & 1.7649 \\ 
        \textbf{} & 2Cr-2F & 1.8056 & 1.7739 \\ 
        \textbf{} & 1Cr-3F & 1.8052 & 1.7651 \\ 
        \textbf{} & 1Cr-4F & 1.9785 & 1.9271 \\ 
        \textbf{} & 2Cr-5F & 1.9811 & 1.9807 \\ 
        \textbf{} & 2Cr-4F & 1.9811 & 1.9807 \\ 
        \textbf{NiF} & 1Ni-1F & 1.7318 & 1.763 \\ 
        \textbf{NiF$_2$} & 1Ni-1F & 1.7424 & 1.7139 \\ 
        \textbf{} & 1Ni-2F & 1.7424 & 1.7139 \\ 
        \textbf{NiF$_3$} & 1Ni-1F & 1.7715 & 1.7373 \\ 
        \textbf{} & 1Ni-2F & 1.7717 & 1.7373 \\ 
        \textbf{} & 1Ni-3F & 1.7688 & 1.7555 \\ \hline
    \end{tabular}
    \label{table_1}
\end{table}

\begin{table}[!ht]
\footnotesize
    \centering
    \caption{Gas-phase valence angles ($\mathrm{^o}$) of various Cr-F and Ni-F structures. Refer to Figure \ref{fig:gases} for the numbering of atoms.}
    \begin{tabular}{p{1cm}p{1.5cm}p{1.1cm}p{1.1cm}}
    \hline
        \textbf{Formula} & \textbf{Three body} & \textbf{ReaxFF} & \textbf{DFT} \\ \hline
        \textbf{CrF$_3$} & 1F-1Cr-2F & 119.8601 & 119.7377 \\ 
        \textbf{} & 1F-1Cr-3F & 120.0521 & 120.1312 \\ 
        \textbf{} & 2F-1Cr-3F & 120.0879 & 120.1312 \\ 
        \textbf{CrF$_4$} & 1F-1Cr-2F & 109.3356 & 109.265 \\ 
        \textbf{} & 1F-1Cr-3F & 109.6039 & 109.6094 \\ 
        \textbf{} & 1F-1Cr-4F & 109.5013 & 109.5428 \\ 
        \textbf{} & 2F-1Cr-3F & 109.6199 & 109.6094 \\ 
        \textbf{} & 2F-1Cr-4F & 109.4932 & 109.5428 \\ 
        \textbf{} & 3F-1Cr-4F & 109.2736 & 109.2581 \\ 
        \textbf{CrF$_5$} & 1F-1Cr-5F & 89.4769 & 89.6005 \\ 
        \textbf{} & 2F-1Cr-5F & 89.5811 & 89.6005 \\ 
        \textbf{} & 3F-1Cr-5F & 90.7739 & 90.7427 \\ 
        \textbf{} & 4F-1Cr-5F & 178.3873 & 178.5145 \\ 
        \textbf{} & 1F-1Cr-2F & 115.4117 & 114.9284 \\ 
        \textbf{} & 1F-1Cr-3F & 122.2318 & 122.5358 \\ 
        \textbf{} & 1F-1Cr-4F & 89.6385 & 89.6005 \\ 
        \textbf{} & 2F-1Cr-3F & 122.3563 & 122.5358 \\ 
        \textbf{} & 2F-1Cr-4F & 89.5806 & 89.6005 \\ 
        \textbf{} & 3F-1Cr-4F & 90.8385 & 90.7427 \\ 
        \textbf{Cr$_2$F$_5$} & 1F-1Cr-5F & 115.3672 & 123.4694 \\ 
        \textbf{} & 3F-1Cr-5F & 114.4 & 120.2549 \\ 
        \textbf{} & 4F-1Cr-5F & 87.8398 & 78.3327 \\ 
        \textbf{} & 1F-1Cr-3F & 108.5097 & 94.127 \\ 
        \textbf{} & 1F-1Cr-4F & 115.3882 & 123.4694 \\ 
        \textbf{} & 3F-1Cr-4F & 114.3772 & 120.2546 \\ 
        \textbf{} & 2F-2Cr-4F & 136.1277 & 142.081 \\ 
        \textbf{} & 2F-2Cr-5F & 136.1375 & 142.081 \\ 
        \textbf{} & 4F-2Cr-5F & 87.6978 & 75.8304 \\ 
        \textbf{NiF$_3$} & 1F-1Ni-2F & 126.1107 & 126.5926 \\ 
        \textbf{} & 1F-1Ni-3F & 116.9482 & 116.7037 \\ 
        \textbf{} & 2F-1Cr-3F & 116.9409 & 116.7037 \\ \hline
    \end{tabular}
    \label{table_2}
\end{table}

\begin{figure}[!ht]
	\centering
	\includegraphics[width=1\textwidth]{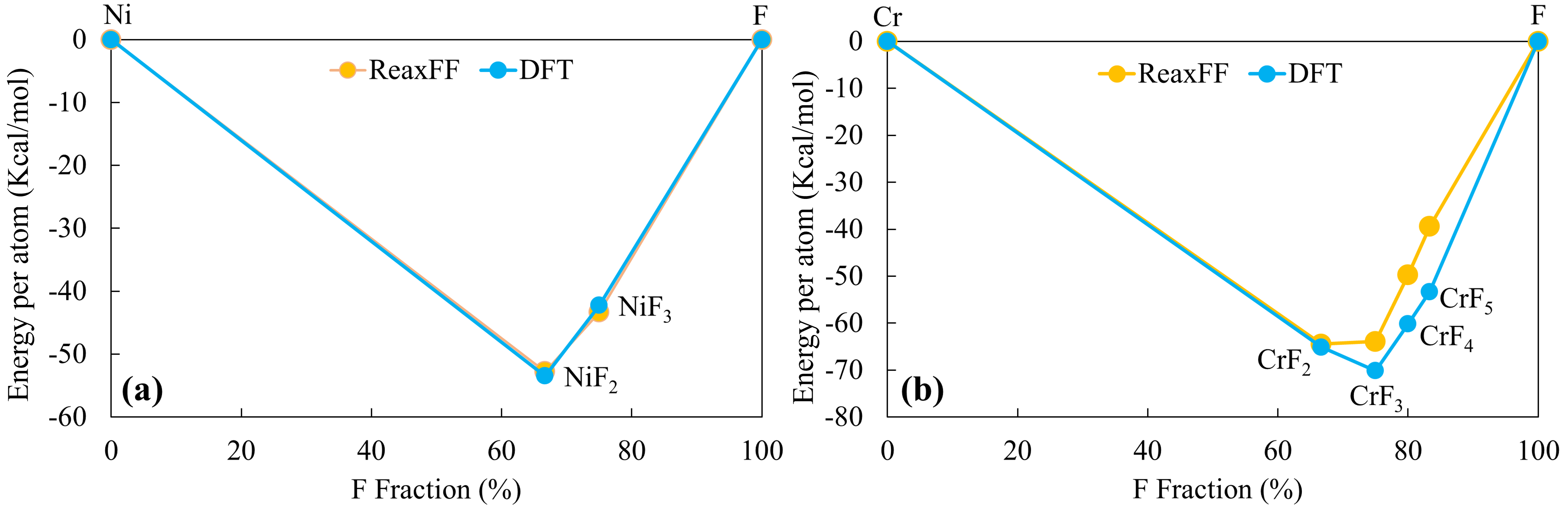}
	\caption{The convex hull plots of the formation energy based on ReaxFF and DFT for different Ni-F (a) and Cr-F (b) crystal structures.}
	\label{fig:formation}
\end{figure}

\begin{figure}[!ht]
	\centering
	\includegraphics[width=1\textwidth]{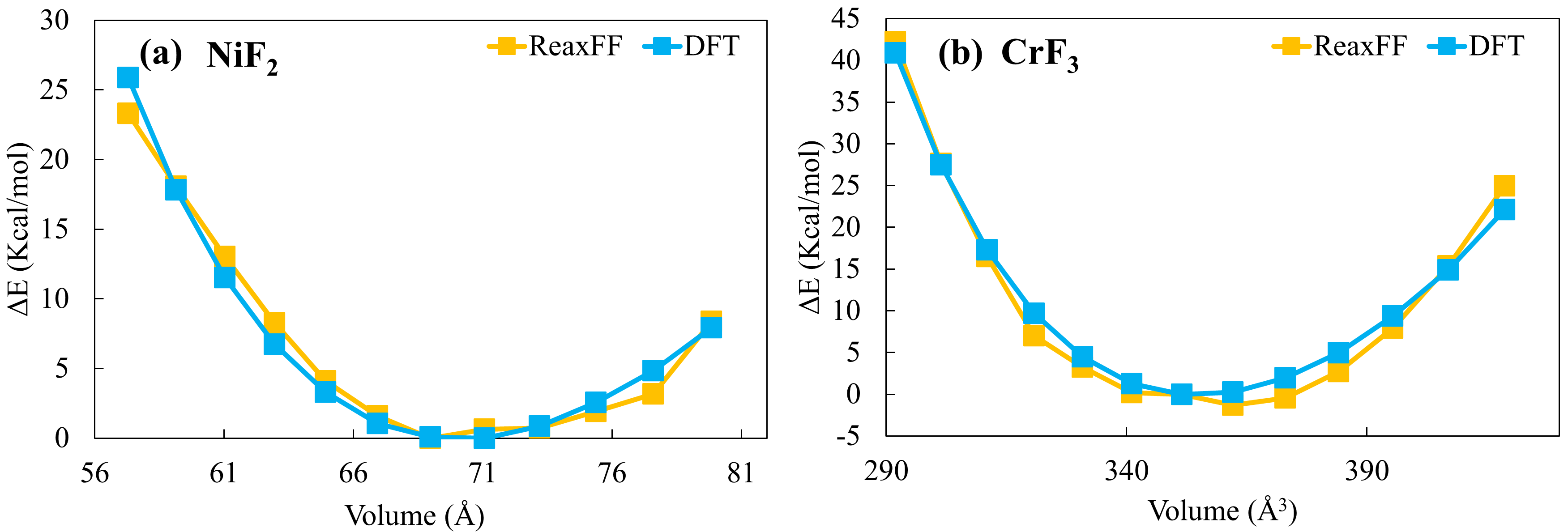}
	\caption{(a-b) The equation of state (EOS) for $\mathrm{NiF_2}$ and $\mathrm{CrF_3}$ from ReaxFF and DFT calculations, respectively.}
	\label{fig:eos}
\end{figure}

To evaluate F-Ni/Cr interaction in condensed forms, Figure \ref{fig:formation} presents a comparative analysis of the ReaxFF and DFT convex hull plots for the formation energy of different Ni-F and Cr-F crystalline structures, extracted from the Materials Project Database \cite{jain2013commentary}. It can be seen that the values derived from ReaxFF generally demonstrate good agreement with the corresponding DFT values. It can be inferred that, $\mathrm{CrF_3}$ (trigonal structure) and $\mathrm{NiF_2}$ (tetragonal structure) exhibit the most negative heat of formation values, being energy favorable structures for Cr-F and Ni-F crystal structures. At the ground state, the lattice constants obtained from ReaxFF are in good agreement with the lattice constants from DFT. Then, we proceed with further calculation of the equations of state (EOS) for $\mathrm{NiF_2}$ and $\mathrm{CrF_3}$ crystalline phases. Figure \ref{fig:eos} displays the comparison across a volume range of -17\% to +20\% of the ground state volume, which highlights the strong agreement between ReaxFF and DFT. By fitting the EOS to the Murnaghan equation of state, the bulk moduli for $\mathrm{CrF_3}$ and $\mathrm{NiF_2}$ are 54 GPa and 109.5 GPa, comparable to literature values of 48 GPa and 109 GPa \cite{jain2013commentary}, respectively. Additional calculation of melting point that reproduces experimental measurement is provided in the SM.

\begin{figure}[!ht]
	\centering
	\includegraphics[width=0.6\textwidth]{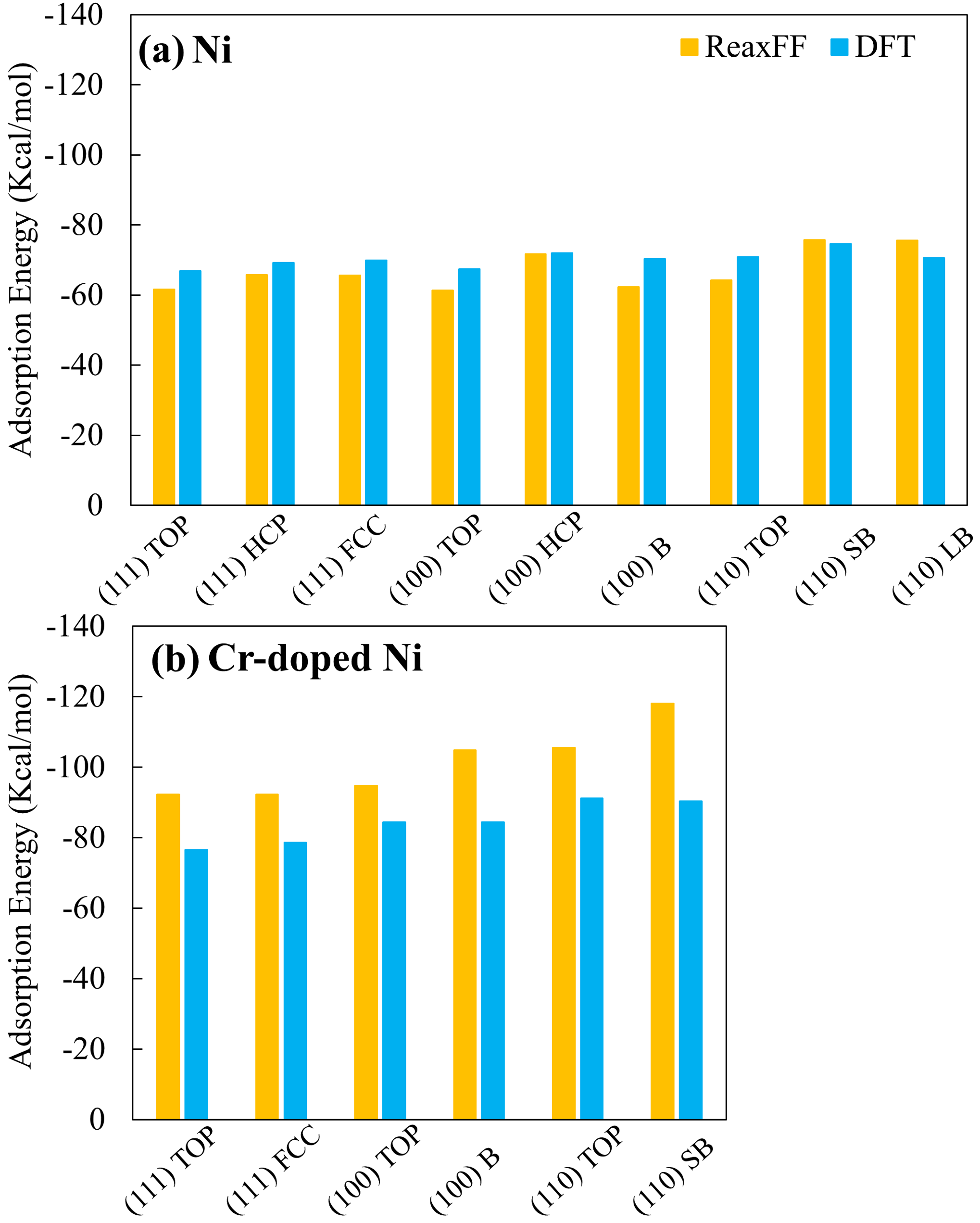}
	\caption{Comparison of F adsorption energies on different surfaces based on ReaxFF and DFT: (a) Ni surfaces with (111), (100), and (110) orientations. (b) Cr-doped Ni surfaces with (111), (100), and (110) orientations.}
	\label{fig:ads}
\end{figure}

F adsorption on three low-index surfaces, i.e., (111), (110), and (100) surfaces are examined for FCC Ni and Cr-doped Ni. The adsorption sites include top, fcc, hcp, bridge (B), hollow, short-bridge (SB), and long-bridge (LB). See SM for detailed structures and the placement of Cr. The DFT calculations yield consistent values with the previous calculations for selected cases by Yin et al. \cite{yin2018theoretical} (see SM). Figure \ref{fig:ads} presents the comparison of the adsorption energies computed by ReaxFF and DFT. The values are in high consistency for Ni surfaces, while ReaxFF tends to overestimate the adsorption energy of F on Cr-doped surfaces.

\begin{figure}[!ht]
	\centering
	\includegraphics[width=0.5\textwidth]{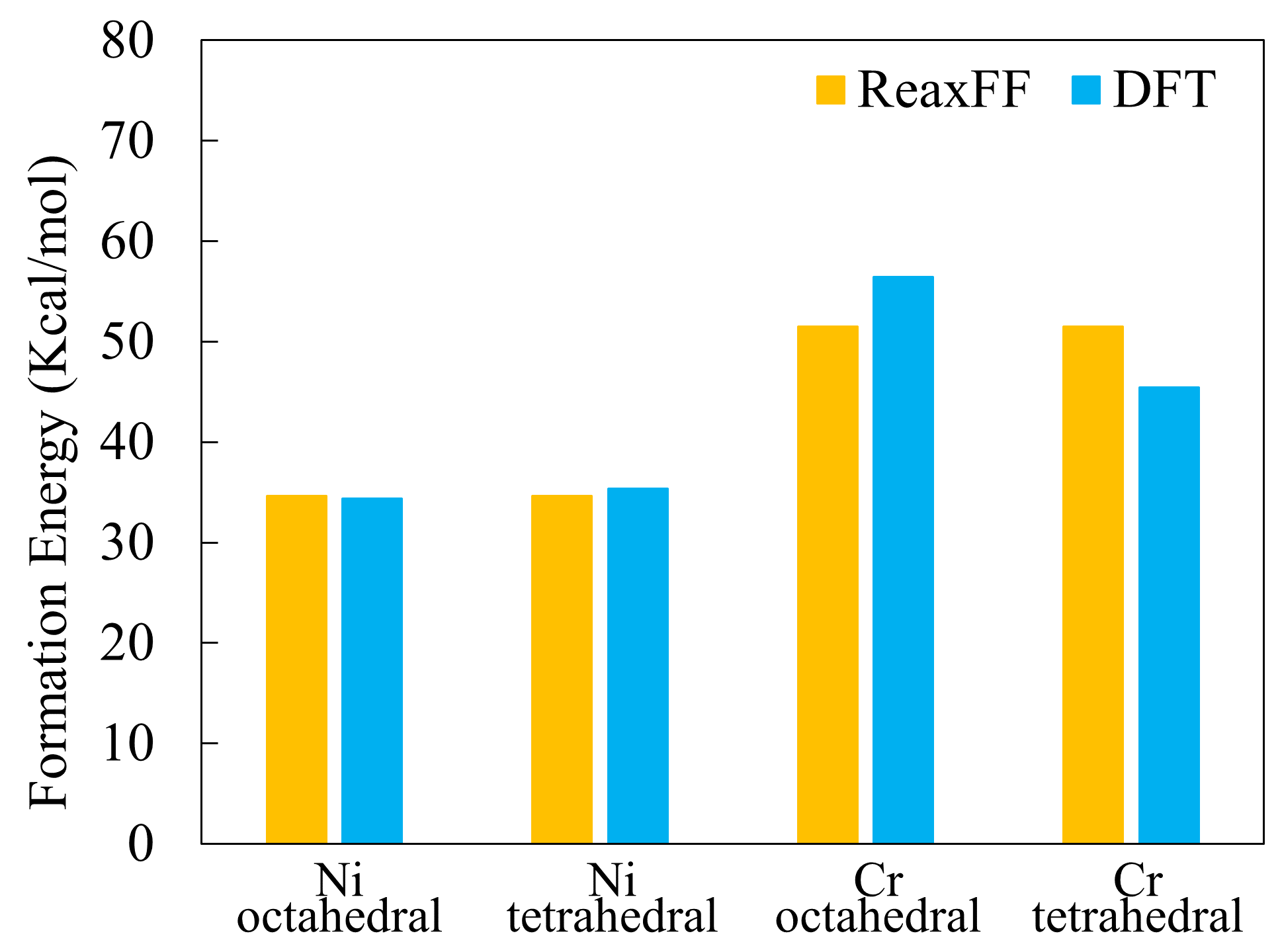}
	\caption{Comparison of F interstitial formation energies at tetrahedral and octahedral sites in FCC Ni and BCC Cr, based on ReaxFF and DFT.}
	\label{fig:inter}
\end{figure}

Finally, F interstitial formation energies at the tetrahedral and octahedral sites, within FCC Ni and BCC Cr metals, are examined. From Figure \ref{fig:inter}, it can be seen that the F incorporation in the metal lattice can be well captured by ReaxFF.

\subsubsection{Li\slash Na\slash K\slash F Force Field}
\begin{figure}[!ht]
	\centering
	\includegraphics[width=1.\textwidth]{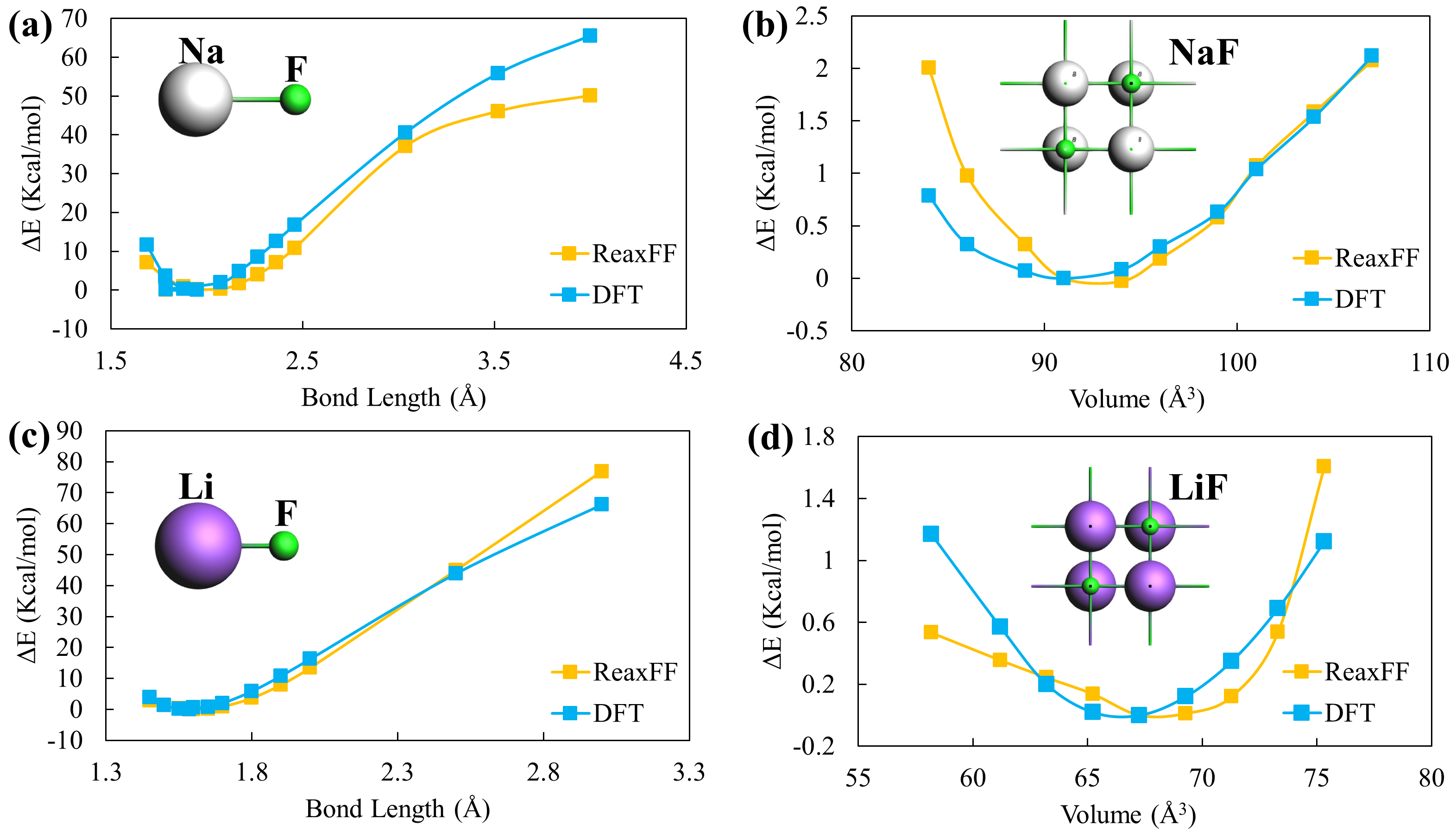}
	\caption{Comparison of EOS curves for (a) NaF molecule, (b) NaF crystal (rock salt), (c) LiF molecule, and (d) LiF crystal (rock salt).}
	\label{fig:LiNa-F}
\end{figure}

To reveal the molten salt structure of FLiNaK, a eutectic mixture of LiF, NaF, and KF, we train the F-Li and F-Na based on the DFT calculated EOS. Note the F-K interactions are obtained from previous work \cite{fedkin2019development}. For validation, we compute the EOS for both NaF and LiF molecules and crystal structures, covering the volume range of -12\% to +12\% of the ground state volume. As shown in Figure \ref{fig:LiNa-F}, values derived from ReaxFF and DFT are in reasonable agreement across the volume range. 

\begin{figure}[!ht]
	\centering
        \captionsetup{width=13cm}
	\includegraphics[width=1.0\textwidth]{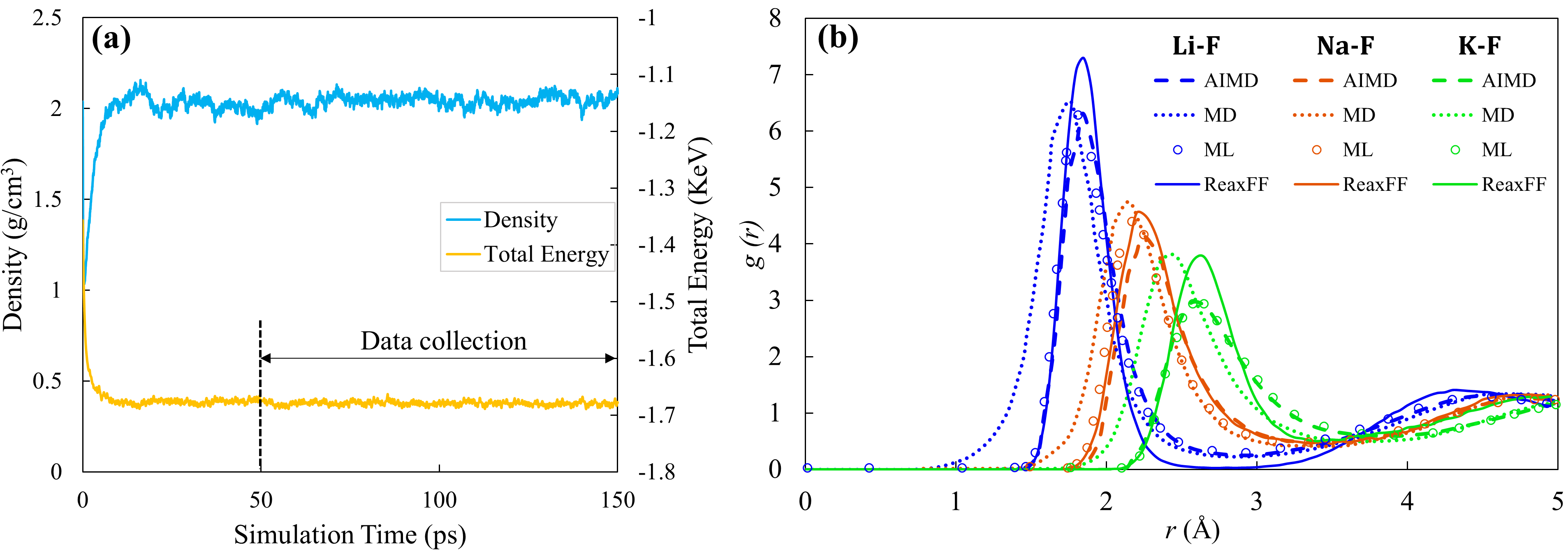}
	\caption{Partial PDFs for Li-F, Na-F, and K-F extracted from ReaxFF MD simulation at 600 $^o$C as compared with other calculated PDFs using AIMD \cite{frandsen2020structure}, machine learning force field \cite{lee2021comparative}, and constant potential MD \cite{langford2022constant}.}
	\label{fig:PDF}
\end{figure}

Additionally, RMD simulations are performed to characterize the salt structure, which can be compared with previous experimental and computational studies \cite{frandsen2020structure,lee2021comparative,langford2022constant,igarashi1988x}. The characteristic partial PDFs are computed for ion pairs Li-F, Na-F, and K-F.  Figure \ref{fig:PDF}(b) shows the time-evolving density and total energy for the simulation setup at 600 $^o$C and ambient pressure. The average density (2.054 g/cm$^3$) stabilizes around the experimentally measured value of 2.035 g/cm$^3$ \cite{frandsen2020structure}. The PDFs are evaluated at the equilibrium stage from 50 ps to 150 ps.

In Figure \ref{fig:PDF}(b), the calculated partial PDFs are compared to results from ab initio molecular dynamics (AIMD) \cite{frandsen2020structure} simulation, constant-potential MD simulation \cite{langford2022constant}, and Machine Learning (ML) force field MD simulation \cite{lee2021comparative}. ReaxFF can well capture the profiles, providing an improved prediction than MD simulations, but less accurate than the trained ML force field, which is associated with a higher computation cost. Moreover, Table \ref{table_3} lists the first peak locations from experiment, AIMD, and ReaxFF, indicating highly consistent results with negligible errors. It confirms that the ReaxFF can yield a reasonable description of the structure of molten FLiNaK. 

\begin{table}[]
\footnotesize
    \centering
    \caption{First peak location for ion pairs determined from X-ray scattering study, AIMD simulation, and RMD simulation at 600 $^o$C.}
    \begin{tabular}{llll}
    \hline
    \multicolumn{4}{c}{First peak (Å)}             \\ \cline{2-4} 
        & Exp. \cite{igarashi1988x} & AIMD \cite{frandsen2020structure} & ReaxFF \\ \cline{2-4} 
    Li$^+$–F$^-$ & 1.83        & 1.84        & 1.85  \\
    Na$^+$–F$^-$ & 2.18        & 2.2         & 2.2   \\
    K$^+$–F$^-$  & 2.59        & 2.6         & 2.65   \\
    F$^-$–F$^-$  & 3.05        & 3.1         & 2.98  \\ \hline
    \end{tabular}
    \label{table_3}
\end{table}

\subsection{The Corrosion Process of NiCr alloys with molten FLiNaK Salt}
\begin{figure}[!ht]
	\centering
	\includegraphics[width=0.4\textwidth]{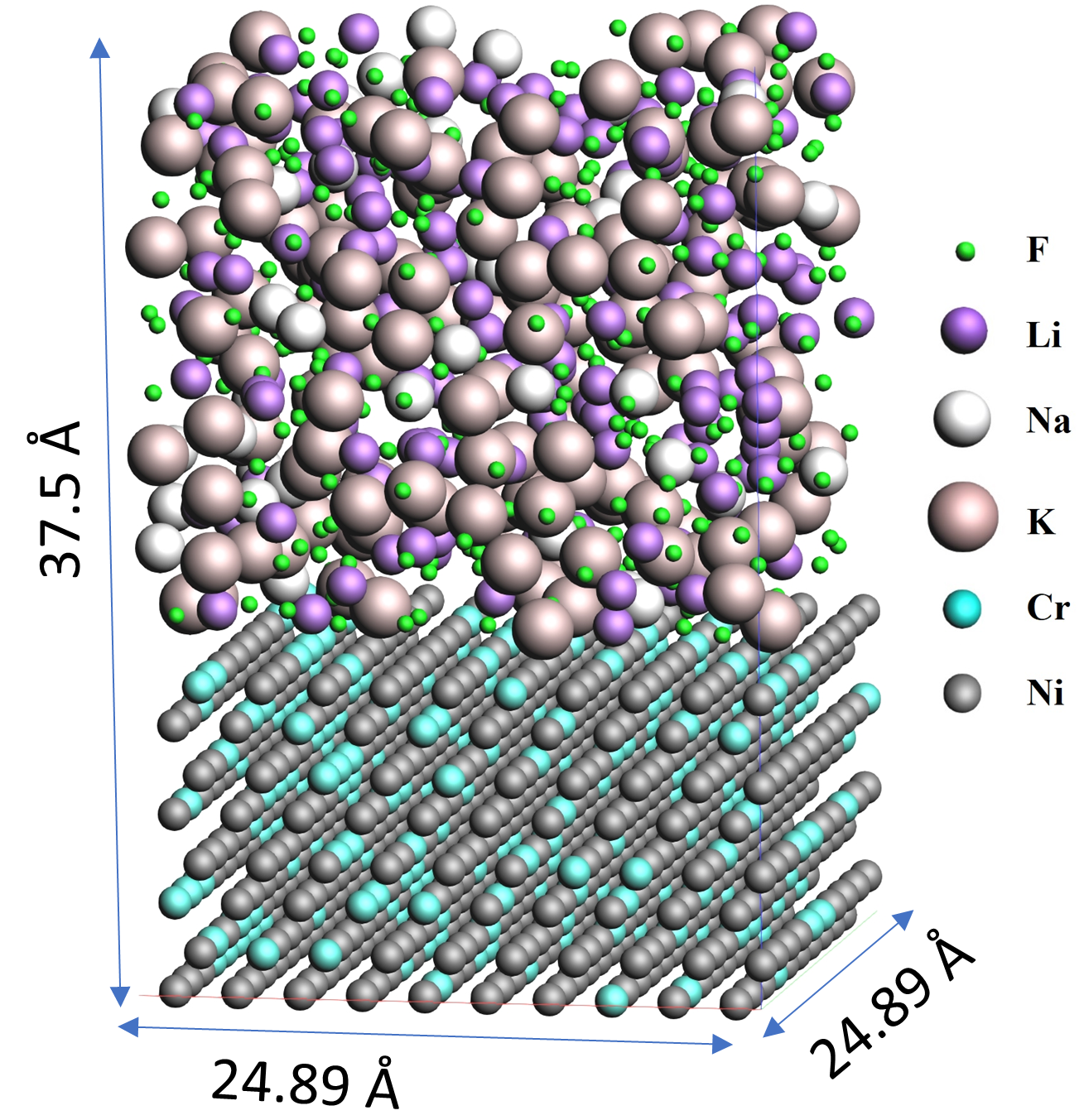}
	\caption{Initial simulation setup: NiCr (100) slab and molten FLiNaK salt.}
	\label{fig:conf}
\end{figure}

To investigate the interfacial behavior of NiCr alloys affected by compositions, RMD simulations are conducted at 1073.15 K (800 $\mathrm{^o}$) for three alloy compositions (Ni$\mathrm{_{0.95}}$Cr$\mathrm{_{0.05}}$, Ni$\mathrm{_{0.85}}$Cr$\mathrm{_{0.15}}$, and Ni$\mathrm{_{0.75}}$Cr$\mathrm{_{0.25}}$). The simulation cell is depicted in Figure \ref{fig:conf}, with a (100) slab of alloy in contact with molten FLiNaK salt, consisting of 163 LiF, 40 NaF, and 147 KF molecules.  

\begin{figure}[!ht]
	\centering
	\includegraphics[width=1.0\textwidth]{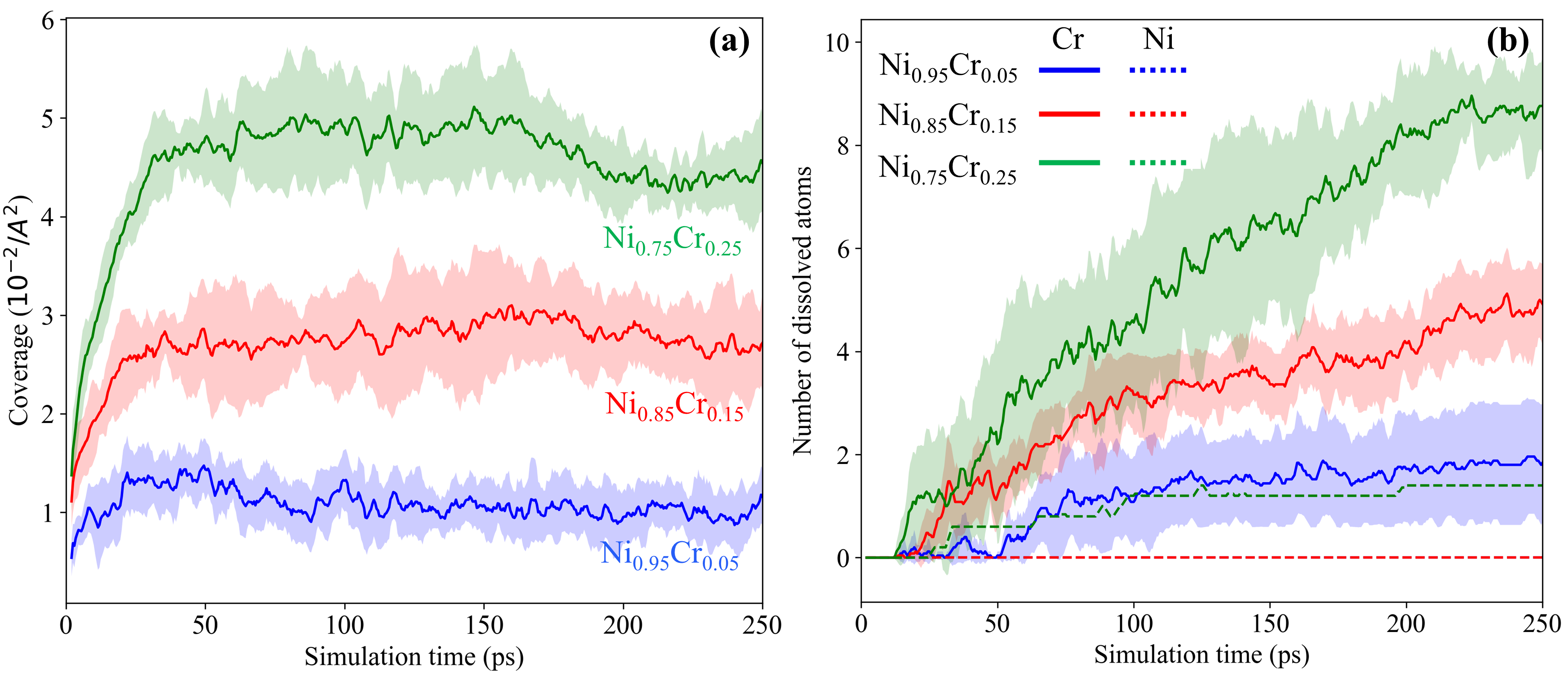}
	\caption{Time evolution of F coverage on the surface of NiCr alloys (a) and the number of metal elements dissolved in the salt (b). Shaded regions indicate the standard deviation.}
	\label{fig:evolution1}
\end{figure}

In all simulations, a general observation includes gradual adsorption of F ions onto the alloy surface, particularly in proximity to Cr atoms, and subsequent dissolution of Cr atoms from the alloy surface into the molten salt. As the local F chemical potential directly contributes to the corrosion rate, we first examine F coverage on the alloy surface. The evolution of F coverage for the various Cr concentrations is depicted in Figure \ref{fig:evolution1}(a). The F coverage serves as an indicator of the interaction between the alloy surface and the salt. With the three alloys, F adsorption quickly approaches a steady state and the saturated coverage increases with Cr concentration. Figure \ref{fig:evolution1}(b) illustrates the number of dissolved metal elements as a function of the simulation time. Cr is mainly dissolved, and a slight dissolution of Ni is observed in the high Cr concentration case. It can be inferred that a higher Cr concentration leads to a high corrosion rate. The dissolution rate slows down with time, even with a high F coverage, which could be attributed to i) decreasing Cr concentration at the surface and ii) the near-surface dissolved metal cations reducing the corrosion potential via interaction with the adsorbed F ions.

Although MD simulations are generally limited by the timescale, this observation of the initial corrosion process is consistent with previous experimental studies: there exists a correlation between weight loss and the Cr content of alloys, with higher Cr content leading to increased corrosion rates \cite{olson2009materials,olson2010intergranular,ouyang2013effect,pillai2021first}. For example, Olson et al. observed that alloys with a lower Cr content, such as Hastelloy N, exhibited less corrosion compared to alloys with a higher Cr content, like Haynes 230 and Inconel 617; Alloy Ni-201, which lacks Cr, is found highly resistant to corrosion \cite{olson2009materials,olson2010intergranular}. Ouyang et al. found that the weight loss is proportional to the original Cr content in several Ni-based alloys (Hastelloy-N, Hastelloy-B3, Haynes-242, Haynes-263) \cite{ouyang2013effect}.  

\begin{figure}[!ht]
	\centering
	\includegraphics[width=1.0\textwidth]{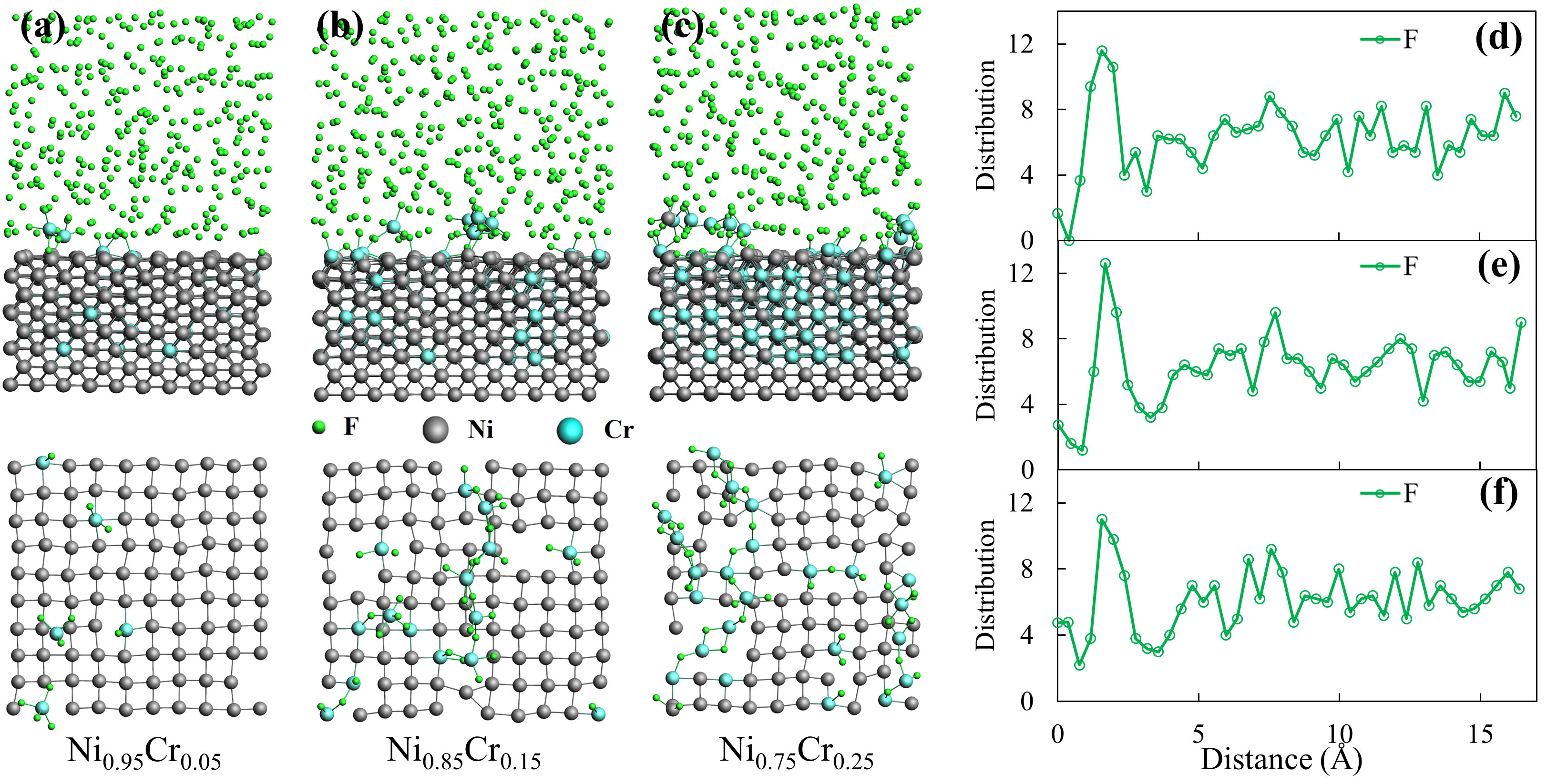}
	\caption{At 250 ps: (a-c) Snapshots for  Ni$\mathrm{_{0.95}}$Cr$\mathrm{_{0.05}}$, Ni$\mathrm{_{0.85}}$Cr$\mathrm{_{0.15}}$, and Ni$\mathrm{_{0.75}}$Cr$\mathrm{_{0.25}}$ alloys, respectively. The bottom row shows the corresponding top view of the metal surface. Salt cations are removed in visualization. (d-f) F distribution versus the distance to the alloy surface for Ni$\mathrm{_{0.95}}$Cr$\mathrm{_{0.05}}$, Ni$\mathrm{_{0.85}}$Cr$\mathrm{_{0.15}}$, and Ni$\mathrm{_{0.75}}$Cr$\mathrm{_{0.25}}$ alloys, respectively.}
	\label{fig:snapshots1}
\end{figure}

To probe the details of the surface changes, Figure \ref{fig:snapshots1} illustrates snapshots of the three different systems at 250 ps. See the complete evolution in supplementary videos. It can be seen that Cr atoms predominately dissolve into the salt, while Ni atoms are generally intact. F ions strongly interact with the surface Cr atoms, facilitating the dissolution process within the timescale. With increasing Cr concentration, F adsorption is enhanced, along with the increase in the number of dissolved Cr atoms. Quantitatively, both the number of dissolved metal elements and adsorbed F ions scale almost linearly with the Cr concentration in the alloy. The distribution of F ions are shown in  Figure \ref{fig:snapshots1}(d-f) for Ni$\mathrm{_{0.95}}$Cr$\mathrm{_{0.05}}$, Ni$\mathrm{_{0.85}}$Cr$\mathrm{_{0.15}}$, and Ni$\mathrm{_{0.75}}$Cr$\mathrm{_{0.25}}$ cases, respectively. The first peak in the distribution corresponds to the F ions that have been adsorbed onto the surface, and the peak positions are independent of the Cr concentration. 

\begin{figure}[!ht]
	\centering
	\includegraphics[width=1.0\textwidth]{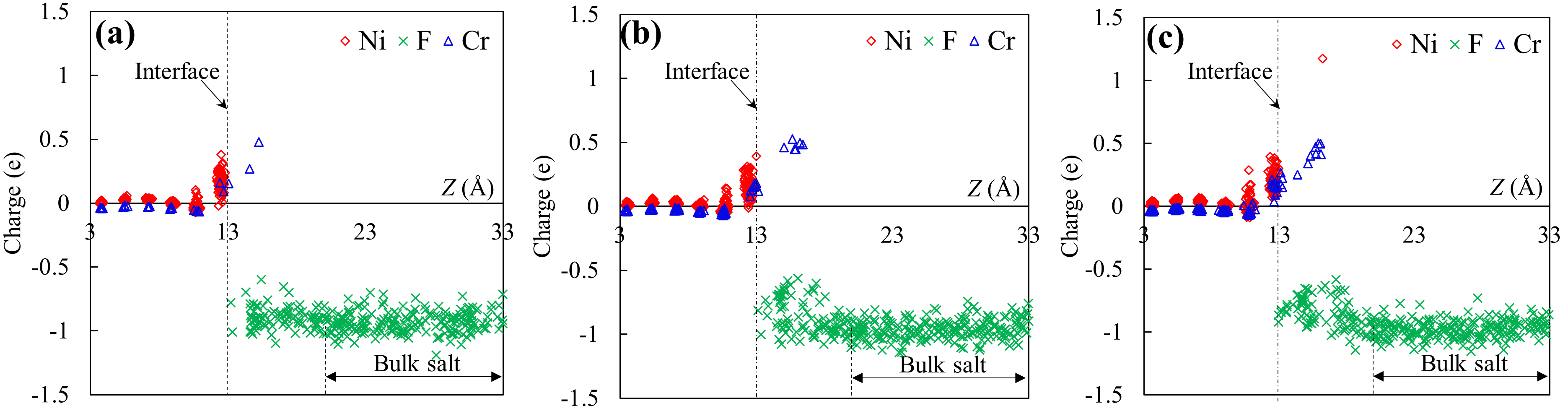}
	\caption{(a-c) Charge distribution at 250 ps for the simulation of Ni$\mathrm{_{0.95}}$Cr$\mathrm{_{0.05}}$, Ni$\mathrm{_{0.85}}$Cr$\mathrm{_{0.15}}$, and Ni$\mathrm{_{0.75}}$Cr$\mathrm{_{0.25}}$ alloys, respectively.}
	\label{fig:charge1}
\end{figure}

Figure \ref{fig:charge1}(a-c) displays the charge distribution of Ni, Cr, and F ions in the three cases at 250 ps, respectively. Within the bulk of the alloy, the atoms maintain a charge close to zero (0 $\pm$ 0.02e), while the major charges occur near the interface due to the affinity with F ions; the charge quickly decays with distance from the surface. The Ni atoms in the bulk are slightly positively charged, and the surface layer has an average charge: +0.158e for Ni$\mathrm{_{0.95}}$Cr$\mathrm{_{0.05}}$, +0.172e for Ni$\mathrm{_{0.85}}$Cr$\mathrm{_{0.15}}$, and +0.206e for Ni$\mathrm{_{0.75}}$Cr$\mathrm{_{0.25}}$ alloys. The average charge for the Ni atoms dissolved in salt (only observed in Ni$\mathrm{_{0.75}}$Cr$\mathrm{_{0.25}}$ alloy case) is +1.39e (Figure. \ref{fig:charge1}c). The bulk Cr atoms are slightly negatively charged in the bulk, and the surface layer has an average charge: +0.128e for Ni$\mathrm{_{0.95}}$Cr$\mathrm{_{0.05}}$,  +0.155e for Ni$\mathrm{_{0.85}}$Cr$\mathrm{_{0.15}}$, and +0.162e for Ni$\mathrm{_{0.75}}$Cr$\mathrm{_{0.25}}$. Upon Cr dissolution, the average charges increase to an average of +0.442e for the three cases. Hence, the average change of charge on average before and after dissolution is +0.314e, +0.287e, and +0.280e for Ni$\mathrm{_{0.95}}$Cr$\mathrm{_{0.05}}$, Ni$\mathrm{_{0.85}}$Cr$\mathrm{_{0.15}}$, and Ni$\mathrm{_{0.75}}$Cr$\mathrm{_{0.25}}$ alloys, respectively. For the F ions, the charge in the bulk fluctuates around -0.95e, consistent across all three cases. Near the interface, an increase of F charge is observed, with an average of -0.901e, -0.829e, and -0.795e within the first peak of PDFs for Ni$\mathrm{_{0.95}}$Cr$\mathrm{_{0.05}}$, Ni$\mathrm{_{0.85}}$Cr$\mathrm{_{0.15}}$, and Ni$\mathrm{_{0.75}}$Cr$\mathrm{_{0.25}}$ cases, respectively. We note that with F adsorption, the average highest charges for Ni, Cr, and F atoms reach +0.35e, +0.24e, and -0.62e, respectively, in close agreement to previous DFT calculations which show the charge difference to be +0.33e, +0.28e, and -0.65e for Ni, Cr, and F atoms upon F adsorption  \cite{ren2016adsorption,yin2018theoretical}.

\subsection{The Corrosion Process of NiCr alloy with individual salt}

\begin{figure}[!ht]
	\centering
	\includegraphics[width=1.0\textwidth]{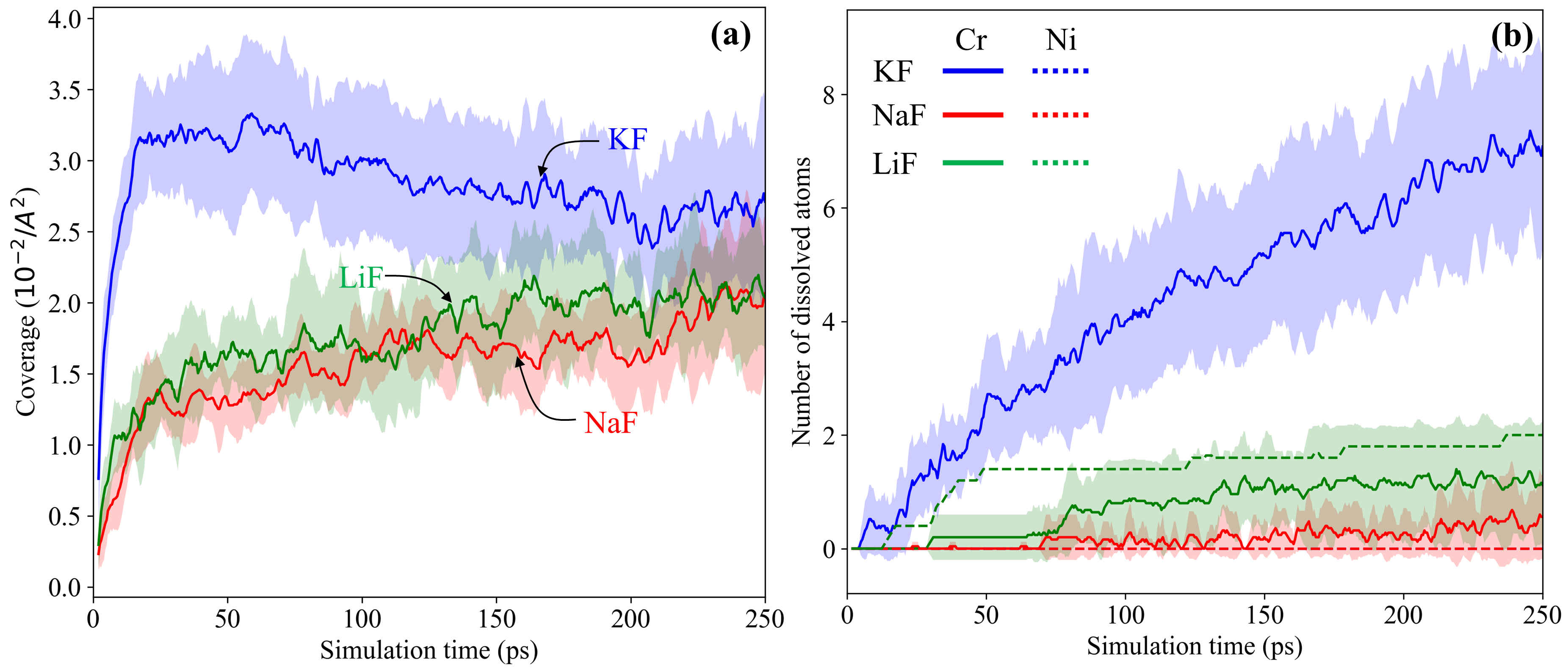}
	\caption{Time evolution of F coverage on the surface of Ni$_\mathrm{0.85}$Cr$_{0.15}$ alloy (a) and the number of metal elements dissolved in the salt (b). Shaded regions indicate the standard deviation.}
	\label{fig:evolution2}
\end{figure}

To identify the effect of individual salt in the molten salt mixture, we also investigated the separate effect of fluoride salt (i.e., LiF, NaF, and KF)  interacting with NiCr alloys at high temperatures. Here, the simulations utilize Ni$_\mathrm{0.85}$Cr$_{0.15}$ slab in the salt environment with  300 salt molecules at 1073.15 K. The molecular number density is consistent across the three salts to avoid additional confounding factors. The evolution of F coverage for the three salt systems is depicted in Figure \ref{fig:evolution2}a, indicating a saturation point that depends on the Cr concentration and salt properties. In KF, a high F coverage is revealed, while NaF and LiF exhibit similar behavior. Reduction in coverage is observed in the case of KF, which is attributed to the fast-reducing Cr concentration on the alloy surface. Figure. \ref{fig:evolution2}b quantifies the dissolved metal atoms versus time. It can be seen that, with KF, the Cr dissolution is strongly increasing with simulation time, while for NaF and LiF, Cr dissolution is significantly lower and remains almost constant across the entire simulation.

\begin{figure}[!ht]
	\centering
	\includegraphics[width=1.0\textwidth]{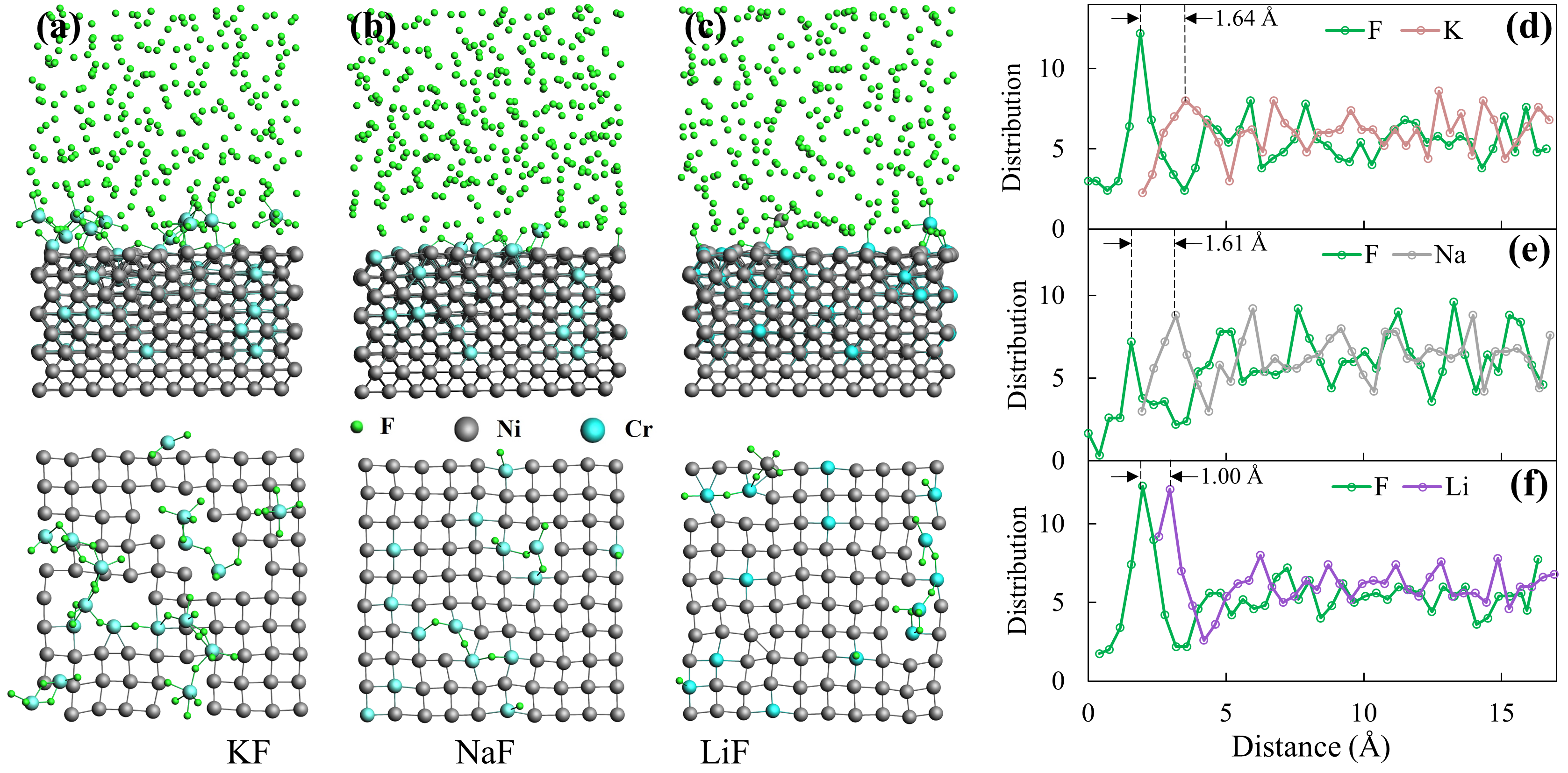}
	\caption{At 250 ps: (a-c) Snapshots of Ni$\mathrm{_{0.85}}$Cr$\mathrm{_{0.15}}$ in KF, NaF, and LiF   respectively. The bottom row shows the corresponding top view of the metal surface. Salt cations are removed in visualization. (d-f) Salt ion distribution versus the distance to the alloy surface for KF, NaF, and LiF, respectively. }
	\label{fig:snapshots2}
\end{figure}

\begin{figure}[!ht]
	\centering
	\includegraphics[width=1.0\textwidth]{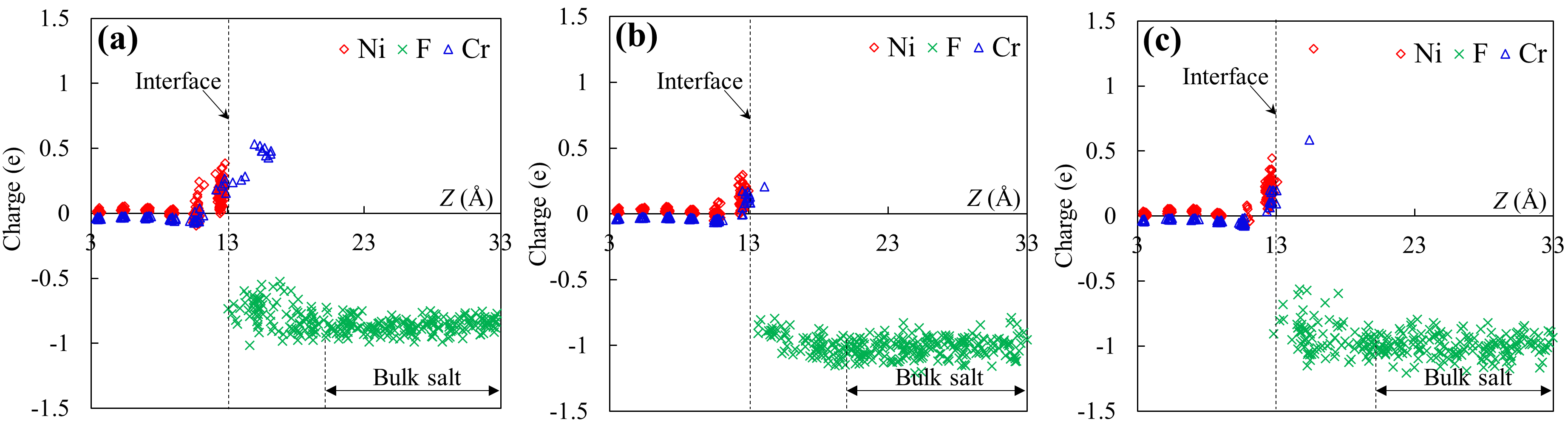}
	\caption{(a-c) Charge distribution at 250 ps for the simulation of KF salt, NaF salt, and LiF salt, respectively.}
	\label{fig:charge2}
\end{figure}

Figure \ref{fig:snapshots2}a highlights the difference in surface modification under the three salt environments after 250 ps. See the complete evolution in supplementary videos. In KF, more metal atoms are dissolved, leading to abundant surface vacancies (Figure \ref{fig:snapshots2}b). Upon metal atom dissolution, we identify various configurations of dissolved Cr/Ni-F complexes formed, including CrF$\mathrm{_2}$, NiF$\mathrm{_2}$, CrF$\mathrm{_3}$, NiF$\mathrm{_3}$, CrF$\mathrm{_4}$, and CrF$\mathrm{_5}$ (see SM Figure. S4). Figure \ref{fig:charge2}(d-f) presents the distribution of ions for the three salts, respectively, which indicate the double layer on the alloy surface. The first peak in F distribution indicates the F ions adsorbed onto the alloy surface. It can be seen that in NaF, the adsorption is weakest as Na and F ions are more dispersed, which reduces the corrosion rate. By comparison, both KF and LiF show increased F adsorption. In KF, the excess adsorption of F ions is not compensated by the K ions, leading to major Cr dissolution. However, in LiF, the first Li peak is close to F peak ($\sim 1.0 \AA$) with similar magnitude due to strong Li-F interaction; this lessens the charge transfer between F and the alloy elements, reducing the corrosion potential. 

The redox potential of the molten salt influences the susceptibility of metals to corrosion \cite{zhang2018redox}. Maintaining a reducing condition in molten fluoride salts can be achieved by the presence of excess amounts of a more active metal or a more stable fluoride-forming metal. Recent experimental results suggest that adding small amounts of Li to the FLiNaK salt can be highly effective in mitigating the selective dissolution of active alloying elements in the structural materials \cite{simpson2012selective,sankar2021effect}, as Li, thermodynamically, possesses the most stable fluoride of all metals engaged in a molten salt corrosion environment. Combining with the current simulations, adding Li metal to the fluoride salt could contribute to a more compact F-Li double layer, reducing the dissolution of alloy species. 

Figure \ref{fig:charge2}(a-c) display the charge distribution of Ni, Cr, and F ions along the z-direction based on one simulation of KF, NaF, and LiF salts at 250 ps, respectively. The patterns are generally consistent with results in Figure \ref{fig:charge1}. Upon Ni dissolution, which only occurs in LiF salt, the charge for the Ni atom increases from +0.192e to +1.306e (a difference of 1.114e). With Cr dissolution, the average change in the charge states is identified: +0.455e, +0.373e, and +0.532e for KF, NaF, and LiF systems, respectively. By comparison, NaF salt leads to the weakest charge transfer under the same ion concentration. Combined with the dissolution behavior (Figure \ref{fig:snapshots2}b), it can be inferred that the corrosion rate in NaF is weakest. In the bulk salt (denoted in Figure \ref{fig:charge2}), the F charge fluctuates around -0.86e, -1.01e, and -0.98e in KF, NaF, and LiF salt, respectively. Near the interface, an increase of F charge is observed, with an average of -0.77e, -0.92e, and -0.88e within the first peak of PDFs for KF, NaF, and LiF salt, respectively.

\section{Conclusion}

In this work, a ReaxFF force field parameter set for Ni/Cr/F/Li/Na/K is developed in order to describe the interactions of the molten FLiNaK salt with NiCr alloys. Training on extensive DFT datasets, the developed force field can well reproduce quantities such as F adsorption energies,  formation energies, and equations of state for various structures. Further validation is accomplished via the dynamic structure of molten FLiNaK where pair distribution functions show good agreement with experimental and first principles calculations.

With this force field, RMD simulations are performed to investigate the interfacial behavior of NiCr alloys at varying Cr-content in molten FLiNaK salt. Consistent with experimental results, molten fluoride salt selectively oxidizes Cr from the interface. Quantitatively, we demonstrate that a high Cr concentration is correlated with a high F coverage on alloy's surface and the number of metal elements dissolved into the salt increases linearly with Cr concentration due to strong Cr-F interactions. To understand the impact of salt composition, RMD simulations are also performed for Ni$_{0.85}$Cr$_{0.15}$ alloy with individual molten salt types (KF, NaF, and LiF). It was found that KF salt leads to the strongest dealloying of Cr atoms from the alloy surface, while NaF and LiF lead to minimal metal element dissolution, with LiF forming a narrow double layer on the alloy surface.

The obtained outcomes highlight the efficacy of the force field for Ni/Cr/F/K/Na/Li in elucidating the intricate interactions between NiCr alloys and molten fluoride salt. Future work would emphasize how the local atomic mechanisms facilitate intragranular and intergranular corrosion under different conditions.

\section{Acknowledgment}
This material is based upon work supported by the U.S. Department of Energy, Office of Science, Office of Basic Energy Sciences Energy Frontier Research Centers program under Award Number DE-SC-0022013.

\bibliographystyle{elsarticle-num} 
\bibliography{science}
\end{document}